\documentclass[aps,prl,reprint,superscriptaddress]{revtex4-1}  

\usepackage{graphicx}

\usepackage{caption}
\usepackage{subcaption}
\usepackage[separate-uncertainty = true]{siunitx}

\usepackage{color}
\usepackage{MnSymbol}

\usepackage{verbatim}
\usepackage[font=small, labelfont=bf]{caption}

\definecolor{color1}{rgb}{0.4,0.4,0.4}
\definecolor{color2}{rgb}{0.4,0.4,0.4}
\definecolor{color3}{rgb}{0.4,0.4,0.4}
\definecolor{color4}{rgb}{1,0,0}
\definecolor{color5}{rgb}{1,0.8,0}
\definecolor{color6}{rgb}{0.4,1,0}
\definecolor{color7}{rgb}{0,1.0,0.4}
\definecolor{color8}{rgb}{0,0.8,1}
\definecolor{color9}{rgb}{0,0,1}
\definecolor{color10}{rgb}{0.8,0,1}

\newcommand{\ket}[1]{|{#1}\rangle}

\newcommand{\braket}[2]{\langle{#1}|{#2}\rangle}

\begin{document}


\title{Ergodicity breaking dynamics of arch collapse }


\author{Carl Merrigan}
\affiliation{Martin Fisher School of Physics, Brandeis University, Waltham MA, 02454}
\email[]{cbrady@brandeis.edu}

\author{Sumit Kumar Birwa}
\affiliation{TIFR International Center for Theoretical Sciences, Shivakote Bengaluru 560089, India}

\author{Shubha Tewari}
\affiliation{Department of Physics University of Massachusetts Amherst,  Amherst MA, 01003}

\author{Bulbul Chakraborty}
\affiliation{Martin Fisher School of Physics, Brandeis University, Waltham MA, 02454}


\date{\today}

\begin{abstract}
Gravity driven flows such as in hoppers and silos are susceptible to clogging due to the formation of arches at the exit whose failure is the key to re-initiation of flow. In vibrated hoppers, clog durations exhibit a broad distribution, which poses  a challenge for  devising efficient unclogging protocols.  
Using numerical simulations, we demonstrate that the dynamics of  arch shapes preceding failure can be modeled as a continuous time random walk (CTRW) with a broad distribution of waiting times, which breaks ergodicity. {Treating arch failure as a first passage process of this random walk, we argue that the distribution of unclogging times is determined by this waiting time distribution. We hypothesize that this is a generic feature  of unclogging, and that specific characteristics, such as  hopper geometry, and mechanical properties of the grains modify the waiting time distribution.}

\end{abstract}


\maketitle


\paragraph{Introduction}

Granular flows are notoriously susceptible to clogging:  the spontaneous arrest of a flow constrained by boundaries and driven towards an opening.  Flows clog due to the formation  of arches, which are structures of mutually stablizing particles spanning the outlet. Understanding the static and dynamic properties of arches is crucial for ensuring smoothly flowing states of grains in silos or pedestrians moving towards an 
exit~\cite{Zuriguel2014,Zuriguel2014b,Helbing2000,Thomas2015}.  Experiments indicate that the distribution of time intervals between clogging events is exponential~\cite{To2001a,Zuriguel2005a,Janda2008a,behringer2009,Kondic2014}. In contrast, the survival times of arches in vibrated silos~\cite{Zuriguel2014b,Lozano2015} or clog durations in intermittent flows~\cite{Janda2009,Mankoc2009a}, exhibit a broad distribution. 
This is {\it a priori} not surprising since arches can have very different geometries and  mechanical stability~\cite{Lozano2012,Hidalgo2013,Lozano2015}. In this work, we show that it is the  {\it dynamical} response of arches to vibrations that leads to the broad distribution of unclogging times.

\begin{figure}[b]
\centering
\begin{subfigure}[h]{0.15\linewidth}
\resizebox{!}{7 cm}{\includegraphics{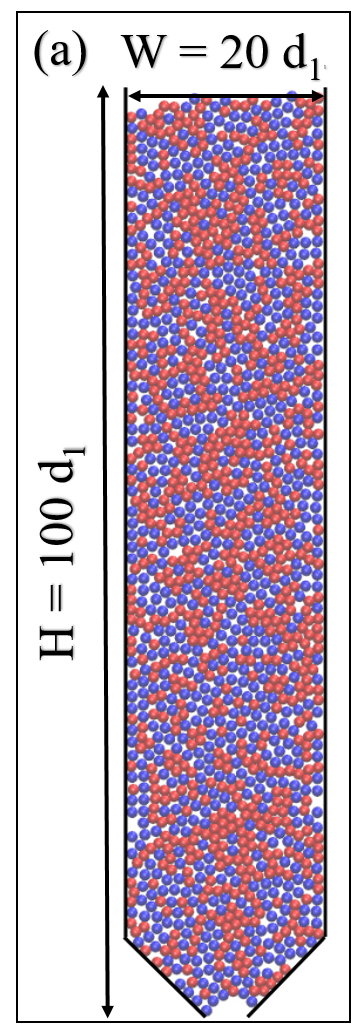}}
\end{subfigure}\hfill
\begin{subfigure}[h]{0.8\linewidth}
\resizebox{0.8\columnwidth}{!}{\includegraphics{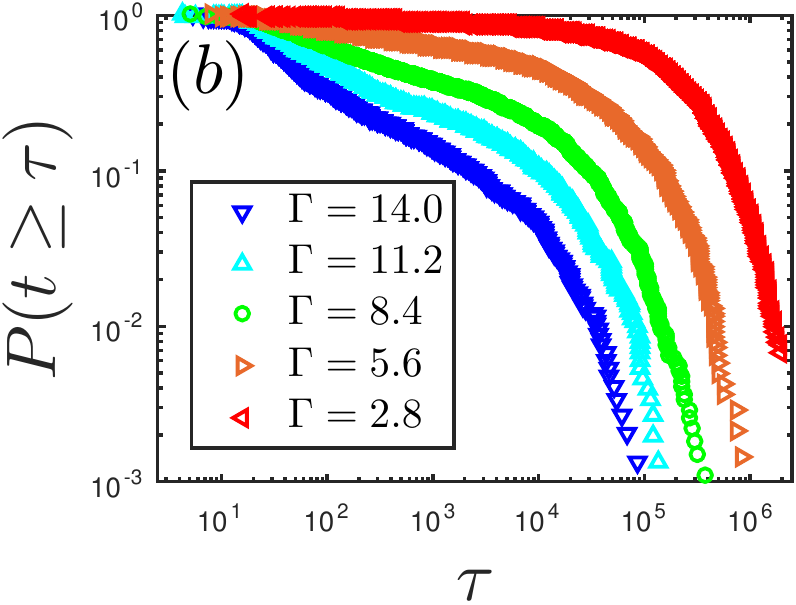}}
\resizebox{0.6\columnwidth}{!}{\includegraphics{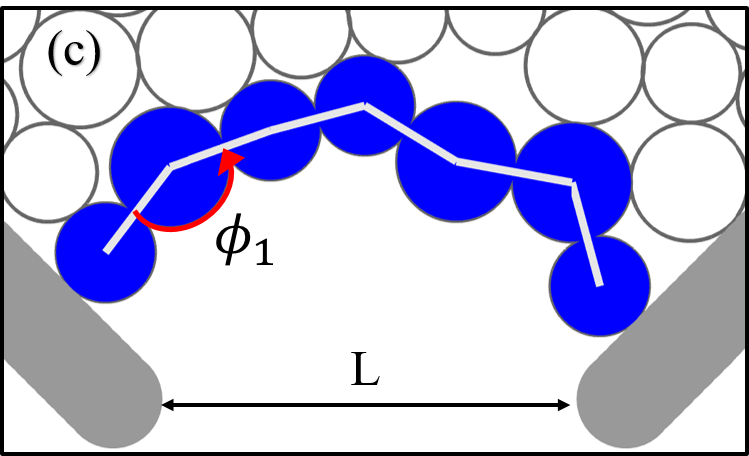}}
\end{subfigure}
\caption{ \label{fig1} (color online) (a) Hopper geometry with width $ W = 20~d_1 $ and height $ H = 100~d_1 $ filled with a bidisperse mixture of $\sim1600 $ spherical grains. (b) Complementary cumulative distribution function $ P(t \geq \tau) $ of the unclogging times for opening size $ L = 4.2~d_1$ and varying vibration amplitudes $ \Gamma $.  (c) Sample five-angle arch with a single opening angle $\phi_1 $ marked. The lower walls are formed from static, closely spaced grains, size $ d_1 $, fixed at $ 45^\circ $ from the horizontal, and are vibrated vertically to unclog the flow.}
\end{figure}

Using Molecular Dynamics simulations~\cite{Plimpton1995,*LAMMPS} of hopper flows, we show that the dynamics of arch shapes are well described by a continuous time random walk  (CTRW) in which the vibrations activate transitions between locally stable arch shapes.  Reminiscent of trap models of the glass transition~\cite{Monthus1996a},  this CTRW   is characterized by a broad distribution of waiting times that leads to ergodicity breaking~\cite{Golding2006,Lubelski2008,He2008b,Jeon2010,Miyaguchi2011,*Miyaguchi2013,Tabei2013}. 

\paragraph{Numerical Simulations}

We perform Molecular Dynamics simulations based on LAMMPS \cite{Plimpton1995,*LAMMPS} using the quasi two-dimensional (2D) hopper geometry shown in Fig~\ref{fig1}.  A   $50-50$ mixture of  bidisperse spheres with diameter ratio $1:1.2$  were randomly distributed within the body of the hopper, allowed to settle under gravity, then flow until a clog develops. We excluded clogged configurations with less than 600 grains remaining in the hopper to ensure a grain depth of at least 1.5 times the hopper width.   The ensemble of clogged states generated using this protocol (further details in~\cite{SI}) were subjected to vibrations to unclog the flow.  The inclined walls at the base of the hopper were displaced vertically at fixed frequency $ f = 10~(g/d_1)^{1/2}$, and varying amplitudes $ A = 1\textendash 5\times 10^{-3}~d_1$.   The vibration strength is characterized by a root-mean-square acceleration, $ \Gamma =\frac{4 \pi^2 f^2 A}{\sqrt{2}}$ that falls in the range of $\Gamma = 2.8\textendash14$ in units of the gravitational acceleration $g$.   The initiation of flow was observed to be caused by arch failures except in rare cases where the arch slides out through the opening before collapsing.

%

The unclogging time  is defined as the time elapsed from the start of the vibrations to the first time the center of any grain exits the outlet.  The  probability distribution function (PDF), $ p(t;\Gamma,L) $, and the complementary cumulative distribution function (CCDF), $ P(\tau;\Gamma,L) = \int_\tau^\infty p(t;\Gamma,L) dt $,  are  estimated from these measurements.    All times, in results presented below,  are quoted in units of the vibration period $ T_{vib} = f^{-1} $.
The CCDF is {estimated} directly by plotting the fraction of all measured unclogging times greater than or equal to each recorded time $ \tau_i $~\cite{Newman2005}.  The distributions were constructed from ensembles with $ N = 1744,1389,2718,1506, {\rm and}~1469 $ different arches for $ \Gamma = 2.8, 5.6, 8.4, 11.2, 14.0 $, respectively. 

The CCDF in Fig.~\ref{fig1}(b) demonstrates that vibrating the hopper produces broad distributions of unclogging times that are  sensitive to the strength of the driving. As the vibration amplitude is reduced, the CCDF becomes broader, indicating an increase in the frequency of arch-breaking events occurring at longer times. The mean unclogging time grows from $ \langle t \rangle = \num{1.58e3} $ for $ \Gamma = 14.0 $, to $ \langle t \rangle = \num{2.5e5} $ for $ \Gamma = 2.8 $. The shape of the CCDF is characterized by  three distinct regions: (i) an initial,  fast decay characterizing  arches that break quickly, (ii) a slower decay and broad plateau region extending over several decades, which crosses over to (iii) another fast decay characterized by a maximum unclogging time. For the smallest amplitude, $ \Gamma = 2.8 $, $ 11 $ out of $ N = 1744 $ arches remained clogged for longer than the maximum simulation time tested $ T_{sim} = \num{2e6} $. Thus the shape of the CCDF can only be estimated up to $ T_{sim} $ for this amplitude. In all other cases, the simulation time was sufficient to break all the arches.   These characteristics are similar to those observed in experiments, except that the experiments report a pure power law tail~\cite{Lozano2015}. 
%
%
\begin{figure}[h]
\begin{subfigure}[t]{0.49\linewidth}
\centering
\resizebox{\linewidth}{!}{\includegraphics{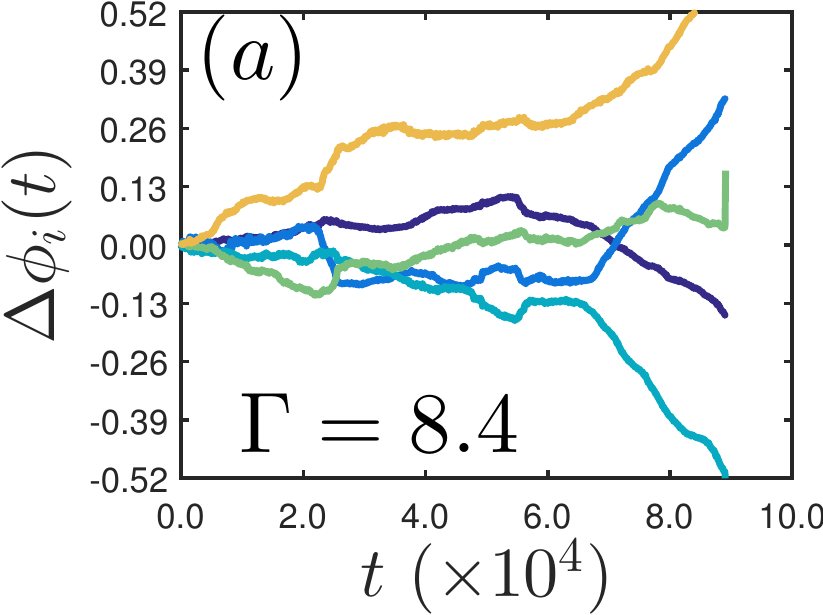}}
\end{subfigure}
\begin{subfigure}[t]{0.49\linewidth}
\centering
\resizebox{\linewidth}{!}{\includegraphics{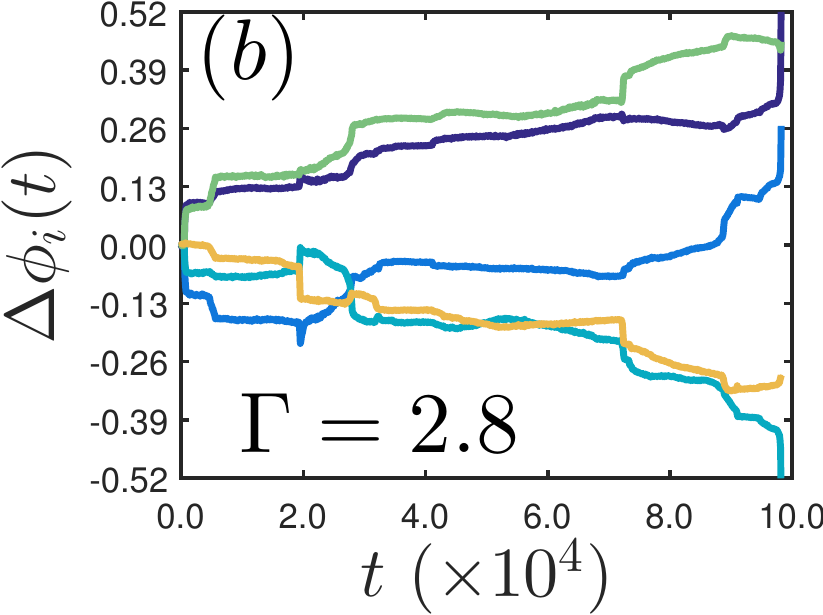}}
\end{subfigure}
\caption{\label{fig2} (color online) Time series of changes to the arch opening angles for two distinct five-angle arches ($ N_\phi = 5 $) unclogged using vibration amplitudes $ \Gamma = 8.4 $ (a) and $ \Gamma = 2.8 $ (b). $ \Delta \phi_i(t) $ indicates the change in radians from each of the initial opening angle values. The unclogging times are $ t = 89,107 $ $ (a) $ and $ t = 98,211 $ $ (b) $. The different angles within the arch (different colors within a panel) show correlated evolution in time. The arches usually undergo a series of reconfigurations, somtimes with changes to the angles as large as $ 30^\circ = 0.52~rad $.}
\end{figure}

\paragraph{Arch Shape Dynamics}

Since the distribution of unclogging times  showed only a weak dependence on the opening size $ L $ for our hopper geometry~\cite{SI} , we analyze  the arch dynamics in detail for a single opening size $ L = 4.2~d_1$.  
The clogging arch is identified as the lowest chain of $N_g$ grains spanning the distance between the outlet walls.  The shape of the arch {\color{blue} } is parameterized by $ N_\phi =  N_g - 2 $ opening angles $ \phi_i(t) $ (Fig.~\ref{fig1}). At $ \L = 4.2 $,  arches with $ N_\phi = 3,4,{\rm and}~5$ dominated the ensemble~\cite{SI}. 

The dynamical response of an arch to the vibration is observed to be a correlated motion of the opening angles. Typical examples of the time evolution of $\phi_i(t)$ are shown in Fig.~\ref{fig2}. (More examples in~\cite{SI}). Characterizing the arch shape by the vector of opening angles $\ket{\phi(t)}= (\phi_1(t), .. ,\phi_{N_\phi}(t))$, we find that $\ket{\phi (t)} $ performs a ``random walk''  in the space of locally stable arch shapes, where each stable shape is characterized by a reconfiguration time that leads to a  ``waiting time'' before the next step in the random walk. The dynamics of the clogging arch can thus be best described by a CTRW with a distribution, $ \psi(t) $, of waiting times. The random environment created by the grains above the arch, including weak and strong force-bearing networks~\cite{behringer2009,Hidalgo2013} is likely responsible for the broad distribution of waiting times. Since a precise definition and direct measurement of these waiting times was not feasible, we infer the features of the waiting time distribution by examining the time and ensemble averages of the mean squared displacement (MSD) of the arch angle vectors. 
  
\begin{figure}[t]
\begin{subfigure}[t]{0.49\linewidth}
\centering
\resizebox{\linewidth}{!}{\includegraphics{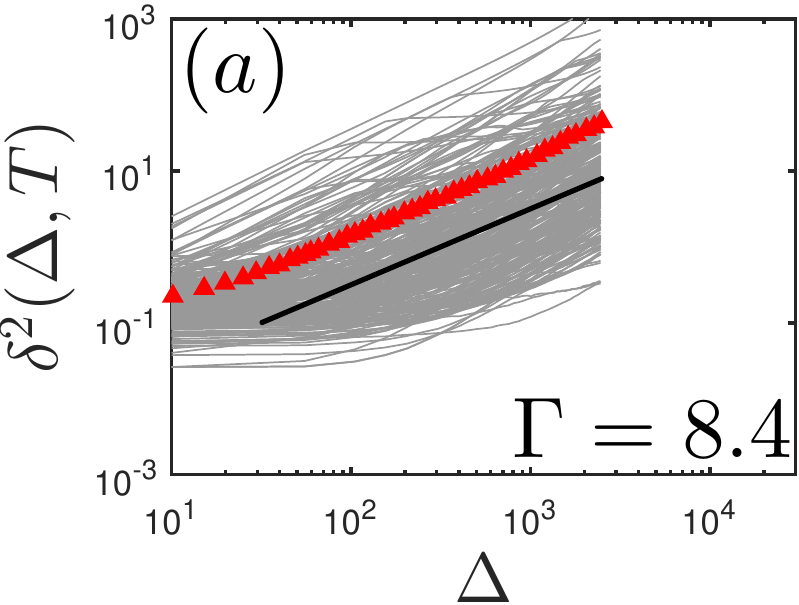}}
\end{subfigure}
\begin{subfigure}[t]{0.49\linewidth}
\centering
\resizebox{\linewidth}{!}{\includegraphics{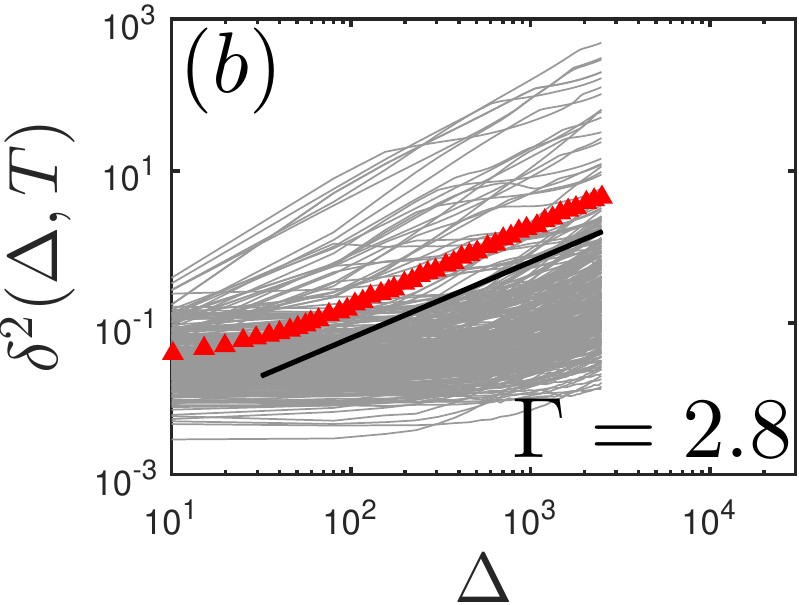}}
\end{subfigure}
\begin{subfigure}[b]{0.49\linewidth}
\centering
\resizebox{\linewidth}{!}{\includegraphics{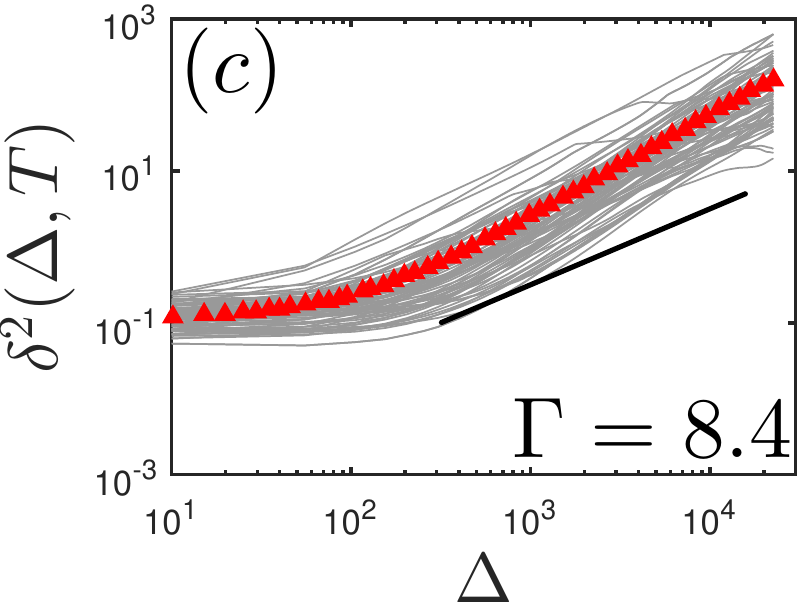}}
\end{subfigure}
\begin{subfigure}[b]{0.49\linewidth}
\centering
\resizebox{\linewidth}{!}{\includegraphics{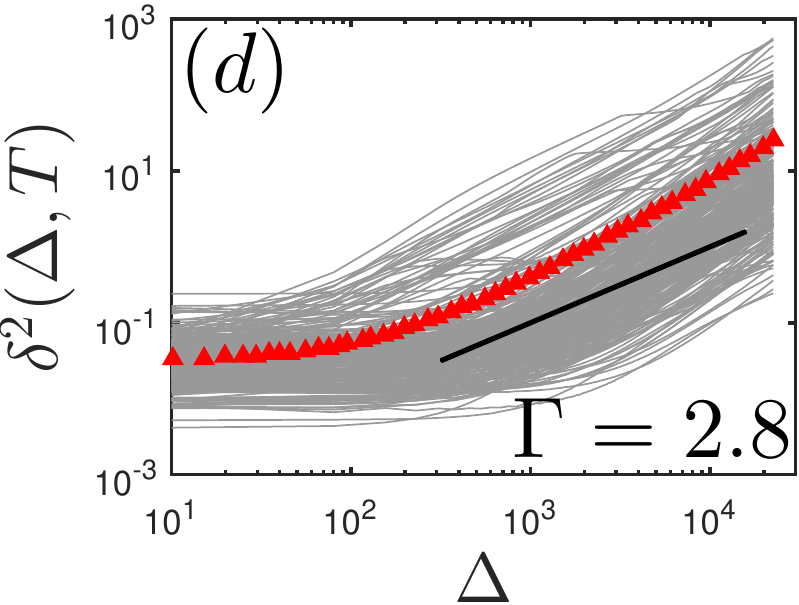}}
\end{subfigure}
\begin{subfigure}[b]{0.49\linewidth}
\centering
\resizebox{\linewidth}{!}{\includegraphics{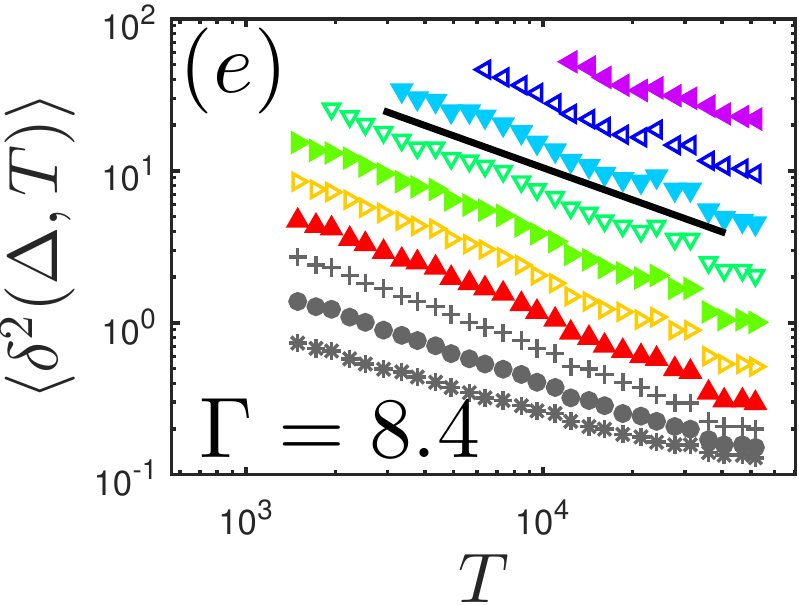}}
\end{subfigure}
\begin{subfigure}[b]{0.49\linewidth}
\centering
\resizebox{\linewidth}{!}{\includegraphics{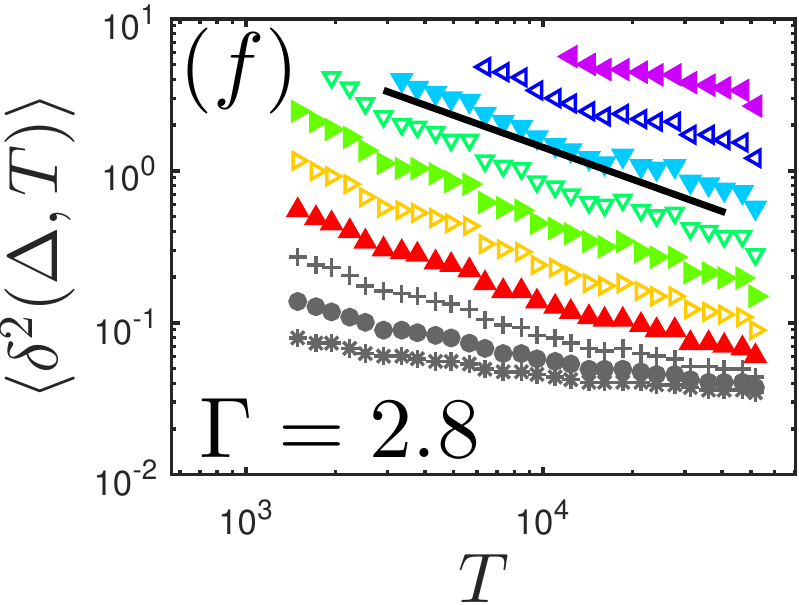}}
\end{subfigure}
\caption{\label{fig3} (color online) (a), (b), (c), (d): Time-averaged mean-squared displacements (TAMSD), $ \delta^2(\Delta,T)$ and their ensemble average,  $ \langle \delta^2(\Delta,T) \rangle $ ({\color{red} $ \filledmedtriangleup $}) shown for two sets of five-angle arches unclogged with vibration amplitudes $ \Gamma = 8.4 $ (left) and $ \Gamma = 2.8 $ (right). The ensembles are contrasted for an intermediate and a long total averaging time, $ T = \num{5e3} $ (a), (b) and $ T = \num{4.5e4} $ (c), (d). The ensembles include all arches that have survived until T or longer. The black lines show a linear slope $ \sim \Delta $, the expected slope for a subdiffusive CTRW with a power law waiting time distribution. (e), (f): Scaling of $ \langle \delta^2(\Delta,T) \rangle $ for both sets of arches. The different symbols show $ \langle \delta^2(\Delta,T) \rangle $ evaluated for fixed values of the lag time $ \Delta = 24 $ ({\color{color1} $ \ast $}),  $ \Delta = 44 $ ({\color{color2} $ \bullet $}),  $ \Delta = 81 $ ({\color{color3} $ \plus $}), $ \Delta = 149 $ ({\color{color4} $ \filledmedtriangleup $}),  $ \Delta = 272 $ ({\color{color5} $ \medtriangleright $}),  $ \Delta = 496 $ ({\color{color6} $ \filledmedtriangleright $}),  $ \Delta = 904 $ ({\color{color7} $ \medtriangledown $}),  $ \Delta = 1649 $ ({\color{color8} $ \filledmedtriangledown $}),  $ \Delta = 3007 $ ({\color{color9} $ \medtriangleleft $}), $ \Delta = 5,484 $ ({\color{color10} $ \filledmedtriangleleft $}). The black line indicates a T dependence $ \sim T^{-0.7 = -1 + 0.3} $.}
\end{figure}

\paragraph{TAMSD and Ergodicity Breaking}
The time-averaged mean-squared displacement (TAMSD) is a standard measure used to characterize random walks\cite{Golding2006,Lubelski2008,Jeon2010,Tabei2013}. For each arch, the TAMSD is defined as 
\begin{equation}
\delta^2(\Delta,T) = \frac{1}{T-\Delta}\int_0^{T-\Delta} \braket{\delta \phi (t,\Delta)}{\delta \phi (t,\Delta)} dt  ~,\nonumber
\end{equation} 
where $\ket{\delta \phi (t,\Delta)}  \equiv \ket{{\phi}(t+\Delta)} - \ket{{\phi}(t)}$,  $ \Delta $ is the lag time, and $ T $ is the total time elapsed since the initiation of vibration. Properties of the underlying stochastic process can be inferred from the behavior of the ensemble-averaged TAMSD, $ \langle \delta^2(\Delta,T) \rangle $, where the angular brackets indicate an average over a set of arches of the same size.  { In a CTRW, the TAMSD are random variables~\cite{He2008b}.  Thus, $ \langle \delta^2(\Delta,T) \rangle $ and  the  ensemble-averaged mean-squared displacement (MSD), without any time averaging, are not necessarily identical. For sets of arches of the same size that are unclogged at the same amplitude, the MSD is calculated as $$ \langle \phi^2(t) \rangle = \frac{1}{N(t)}\sum_{i = 1}^{i=N(t)} \Delta \vec{\phi}_i(t)\cdot \Delta \vec{\phi}_i(t)  ~,$$ where the index $ i $ indicates individual arches~\cite{SI}. 

For a simple random walk,  the $\delta^2(\Delta,T)$ are narrowly distributed around their mean $ \langle \delta^2(\Delta,T) \rangle $, and for large T, the TAMSD for a single walker approaches the ensemble-averaged MSD: $ \delta^2(\Delta,T \rightarrow \infty) = \langle \phi^2(t = \Delta) \rangle $.   The class of CTRWs, characterized by a power law, $ \psi(t) = t^{-\alpha + 1}~, t\rightarrow \infty$, with  $ 0 < \alpha < 1$, is known to break ergodicity with time averages being different from ensemble averages~\cite{Lubelski2008,He2008b}.  $ \langle \phi^2(t) \rangle$  is subdiffusive with an anomalous exponent $\alpha$: $\langle \phi^2(t) \rangle \propto t^{\alpha}$ (see~\cite{klafter2011}). However, $ \langle \delta^2(\Delta,T)$ is diffusive in $\Delta$, with  the scaling form $ \langle \delta^2(\Delta,T) \rangle \sim \frac{\Delta}{T^{1 - \alpha}} $ $ (\Delta << T) $~\cite{Lubelski2008,He2008b,klafter2011,Miyaguchi2011,*Miyaguchi2013}. We observe both a subdiffusive  MSD~\cite{SI}, and a diffusive growth of $ \langle \delta^2(\Delta,T) \rangle $ in the arch dynamics.  }

Fig.~\ref{fig3} compares many $ \delta^2(\Delta,T)$ and their  ensemble average,  $ \langle \delta^2(\Delta,T) \rangle $
for two different sets of five-angle arches ($ N_{\phi} = 5 $) subjected to vibration amplitudes $ \Gamma = 8.4 $ and $ \Gamma = 2.8 $. These amplitudes were chosen for the main analysis of the arch dynamics because their $ p(t;\Gamma,L) $ differ significantly.  In addition,  there are a  sufficient number of long-lived arches to provide adequate statistics for time and ensemble averaging~\cite{SI}. For  $ T = 5000 $, there is a broad scatter in $ \delta^2(\Delta,T)$  around $ \langle \delta^2(\Delta,T) \rangle $ at both values of $\Gamma$. This broad scatter  is a clear signature of ergodicity breaking~\cite{Lubelski2008,Jeon2010,RgBewerunge2016a}. For the longer averaging time $ T = 45000 $, the broad scatter is still clearly present at $ \Gamma = 2.8 $. At  $ \Gamma = 8.4 $, however,  there is an apparent narrowing of the distribution, which hints at a possible recovery of ergodicity for long enough trajectories. We will address this feature below in the context of the precise functional form of the waiting time distribution $\psi(t)$ and its connection to the resulting unclogging time distribution in Fig~\ref{fig5}. It is also clear that the $\Delta$ scaling of $ \langle \delta^2(\Delta,T) \rangle $ emerges only for $\Delta > \Delta_0 \approx 10^2$. The CTRW model is thus able to capture the dynamics of the arches at time scales much longer than a vibration period.

In Fig.~\ref{fig3} $(e)$ and $(f)$, we plot $ \langle \delta^2(\Delta,T) \rangle $ as a function of $T$ for five-angle arches at  $ \Gamma = 8.4 $ and $ \Gamma = 2.8 $,  at different values of $\Delta$. As seen from the figure, $\langle \delta^2(\Delta,T) \rangle$  follows the predicted scaling form $ \langle \delta^2(\Delta,T) \rangle \sim \frac{\Delta}{T^{1 - \alpha}} $  with $ \alpha  \simeq 0.3 $ for the range $\Delta_0 < \Delta << T$. Within statistical errors, results for both the amplitudes were consistent with this exponent. Comparing the two amplitudes (vertical axes in Fig.~\ref{fig3} $(e)$ and $(f)$), $ \langle \delta^2(\Delta,T) \rangle $ shows a change in magnitude of $ \approx 10 $, which indicates a decrease in the magnitude of the effective diffusion coefficient $ D_\alpha $ as the amplitude decreases from $\Gamma=8.4$ to $\Gamma=2.8$. The same reduction in magnitude is also present in the ensemble-averaged MSD $\langle \phi^2(t) \rangle$~\cite{SI}
    
\begin{figure}

\begin{subfigure}[t]{\linewidth}
\centering
\resizebox{!}{5cm}{\includegraphics{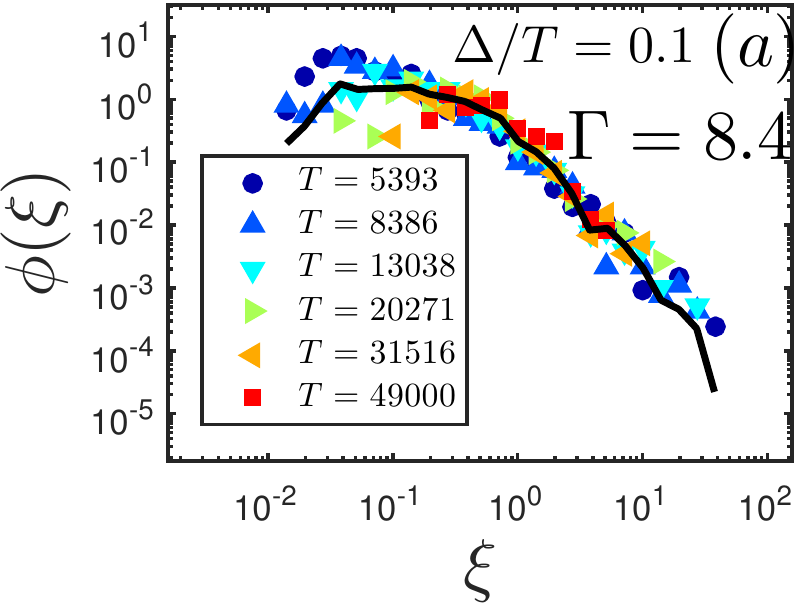}}
\end{subfigure}
\begin{subfigure}[t]{\linewidth}
\centering
\resizebox{!}{5cm}{\includegraphics{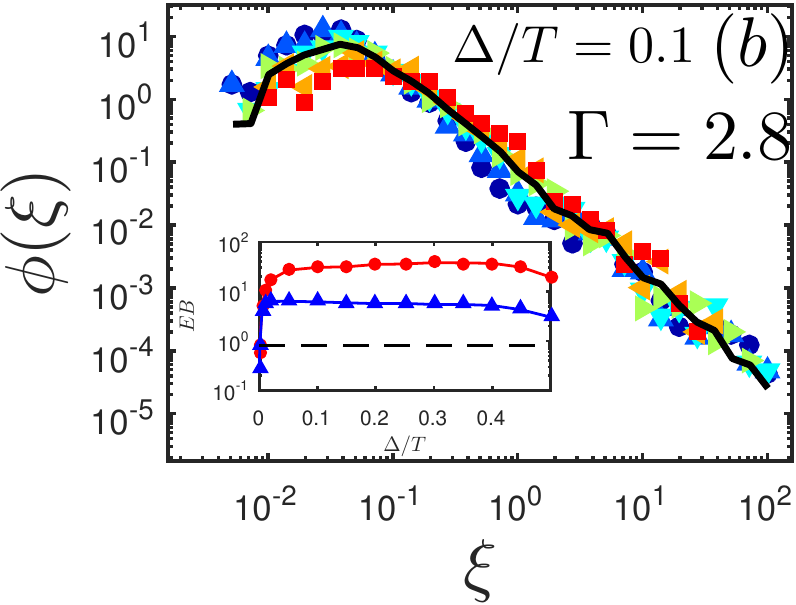}}
\end{subfigure}
\caption{\label{fig4}(color online) Distributions $ \phi(\xi) $ of the normalized TAMSD $ \xi = \frac{\delta^2}{\langle\delta^2\rangle} $ for $ \Gamma = 8.4 $ $ (a) $ and for $ \Gamma = 2.8 $ $(b)$. The shapes of the distributions were found to depend on the ratio $ \Delta/T $, and are shown here for $ \Delta/T = 0.1 $ and different $ T $ values given in the legend, which are the same for both plots. The averaged $ \phi(\xi) $, black solid line,  is used to compute the EB at $\Delta/T=0.1$.  Inset to (b): EB parameter, $ \langle \xi^2 \rangle - \langle \xi \rangle^2 $ for amplitudes $ \Gamma = 2.8 $ ({\color{red} $ \bullet $}) and $ \Gamma = 8.4 $ ({\color{blue} $ \filledmedtriangleup $}) as a function of $ \Delta/T $, computed using averaged $\phi(\xi)$ over a range of $\Delta/T$~\cite{SI}. The EB parameter is roughly constant in the range $ \Delta/T = 0.1\textendash0.4 $. The dashed black line indicates the expected EB value for a power law $\psi(t)$ with $ \alpha = 0.3 $. }
\end{figure} 

\begin{figure}[b]

\begin{subfigure}[t]{\linewidth}
\centering
\resizebox{!}{5.5 cm}{\includegraphics{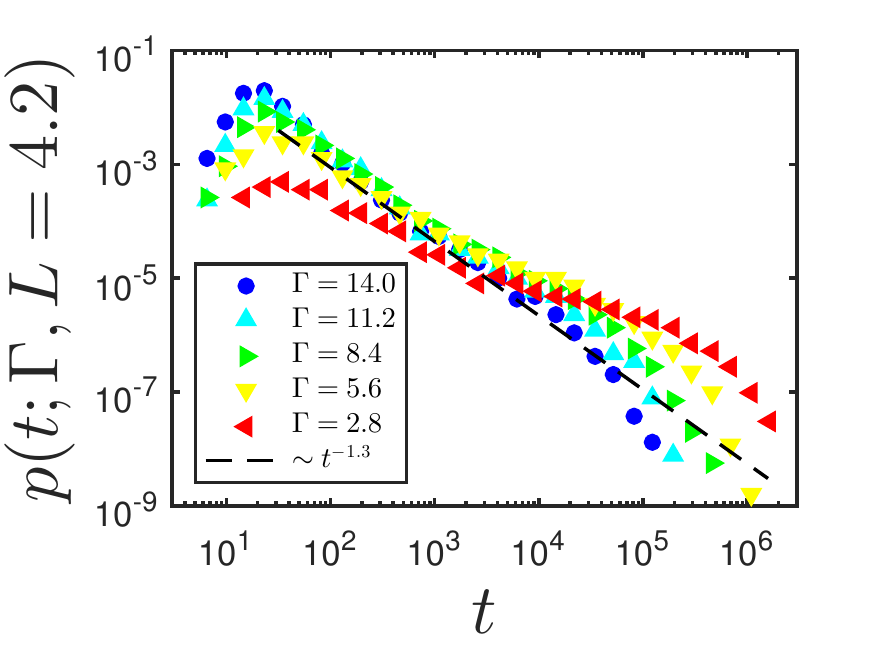}}
\end{subfigure}
\caption{\label{fig5}(color online) Probability distribution functions $ p(t;\Gamma,L)) $ for the unclogging times measured for varying vibration strengths $\Gamma =  2.8 , 5.6 , 8.4 , 11.2, 14.0  $. These distributions are seen to have a peak at short unclogging times, a broad, similar power-like like decay over an intermediate range of times, and finally an upper cutoff time scale at long times. The tails of the distribution show an increasingly dominant plateau, indicating the weight in the unclogging time distributions is being redistributed from shorter to longer unclogging times.} 
\end{figure} 

\paragraph{Effects of vibration amplitude} 

A CTRW differs crucially from a simple random walk in that the $ \delta^2(\Delta,T)$ are  random variables~\cite{Lubelski2008,He2008b,klafter2011}. The properties of this stochastic variable are characterized  by the distribution $\phi(\xi)$ of the scaled variable 
$\xi (\Delta, T)= \delta^2 (\Delta,T)/\langle\delta^2\rangle (\Delta,T)$.
For a pure random walk, this distribution approaches a delta function $\delta(\xi -1)$ at large $T$. In contrast, for a CTRW with a power law distribution of waiting times, $\phi(\xi)$ approaches a universal form parametrized by the power-law exponent $\alpha$ at large $\Delta$ and $T$~\cite{He2008b}.  The form of $\phi(\xi)$ provides  information about the underlying stochastic process beyond that contained in the ensemble average $ \langle \delta^2(\Delta,T) \rangle $. In particular,  the variance $\rm{EB} \equiv \langle \xi^2 \rangle - \langle \xi \rangle^2 $ provides a {\it quantitative} measure of ergodicity breaking~\cite{He2008b}.

We measured the distribution $ \phi(\xi) $ for the same ensemble of arches that were considered in the TAMSD study.  For  $\Gamma=8.4 ~\rm{and}~ 2.8$, we find that $ \phi(\xi) $ depends only on the ratio $ \Delta/T $~\cite{SI}. In Fig. \ref{fig4}, we plot $\phi(\xi)$, and EB, obtained by averaging over a large range of both $\Delta$ and $T$, as a function of $ \Delta/T $. For $ \Delta/T = 0.1\textendash0.4 $, the form of the distribution and its variance are roughly independent of $\Delta/T$. Remarkably, this value depends sensitively on $\Gamma$, with an average value of $ 40 $ for $ \Gamma = 8.4 $ and $ 6 $ for $ \Gamma = 2.8 $. Both these values are much larger than the asymptotic prediction ($ EB = 0.8$) for a pure power law $\psi(t)$ with $ \alpha = 0.3 $~\cite{He2008b}. 
It is clear that $\phi(\xi)$ for the arch dynamics has significantly more weight at  $\xi >> 1$ than expected from  a power-law CTRW, and a concomitant shift in the peak to $\xi <1$ to maintain a mean of unity.
  
Both $ \phi(\xi) $ and the unclogging time distribution $ p(t;\Gamma,L) $ should be sensitive indicators of the form of $ \psi(t) $ governing the transitions between locally stable arch structures. Since $ \phi(\xi) $ indicates  a larger sample-to-sample heterogeneity of the arch dynamics than  expected for a power-law CTRW consistent with the scaling of $ \langle \delta^2(\Delta,T) \rangle $, $ p(t;\Gamma,L) $ should also show a corresponding deviation from a pure power law behavior. As seen in Fig~\ref{fig5},  the distributions $ p(t;\Gamma,L) $ are indeed broader than that expected from $\psi(t) \propto t^{-\alpha +1}$  with $\alpha = 0.3$. As $\Gamma$ is reduced, weight  in the distribution is shifted from short times to long times, consistent with the changes  of $\phi(\xi)$ with $\Gamma$. We can, therefore, conclude that (a) the waiting-time distribution characterizing the arch dynamics is broader than a power-law, and (b) as the vibration amplitude is decreased, the ensemble of arches becomes increasingly dominated by ones that live longer than expected from a power-law. We note, however, that there is a clear cutoff in the unclogging time distribution that grows as $\Gamma$ is reduced (CCDF in Fig. \ref{fig1}). 

The distribution of first passage times of a CTRW has an upper cutoff only if it occurs in a bounded volume~\cite{redner2001} and if, in addition, $\psi(t)$ has a large timescale cutoff~\cite{Metzler2004,Miyaguchi2011,*Miyaguchi2013}.  The space of arch-angles is indeed bounded. For any finite system, a cutoff in $\psi(t)$ is also to be expected since arches will ultimately break or simply fall out of the hopper. Finite-size effects have been studied in detail for power-law CTRWs\cite{Miyaguchi2011,*Miyaguchi2013}. In experiments, the unclogging time distributions\cite{Zuriguel2014b,Lozano2015} do not seem to have a characteristic cutoff time, which possibly reflects the fact that this time is much larger than the duration of the experiment.

\paragraph{Conclusions}

Detailed analysis of the dynamics of arch shapes in response to vibrations demonstrates that clogging arches break ergodicity.  The arches evolve in a landscape of locally stable shapes reminiscent of trap models~\cite{Monthus1996a} with anomalously broad distribution of trap depths.  Mapping the dynamics to a CTRW, we quantify the degree of ergodicity breaking and show that it increases with decreasing vibration amplitude.   This mechanism explains the broad distribution of unclogging times observed in our numerical simulations, and experiments~\cite{Zuriguel2014b,Lozano2015}. We find that the distributions are broader than expected from a power-law distribution of waiting times, which in turn follows from an exponential distribution of trap depths~\cite{Monthus1996a}.  Recent analysis of experiments~\cite{Zuriguel2017} in terms of a trap model indicates that the distribution of trap depths is a stretched exponential.   Such a distribution could lead to the strong ergodicity breaking observed in the simulations.   It would be interesting to measure the extent of ergodicity breaking in the vibrated-hopper experiments.




%




{\it Acknowledgements:}
The work of CM and BC has been supported by NSF-DMR 1409093 as well as the Brandeis IGERT program. 
SKB thanks Narayanan Menon and Rama Govindarajan for their support, and acknowledges funding from TCIS Hyderabad, ICTS Bangalore, the
APS-IUSSTF for a travel grant and NSF-DMR 120778 and NSF-DMR 1506750 at UMass Amherst.
We would like to thank Iker Zuriguel, {\'{A}}ngel Garcimart{\'{i}}n, Aparna Baskaran, and Kabir Ramola for helpful discussions and comments. 

%

\end{document}



\title{Supplementary Information: Ergodicity breaking dynamics of arch collapse}








\date{\today}

\begin{abstract}

Here were offer additional information on the implementation of the simulations in LAMMPS, including the construction of the hopper geometry, the material parameters used for modeling the grains, and the protocol for the creation of the clogged states. We also include more examples of arch angle time series in order to illustrate the nature of the arch dynamics. The averaging procedures for the regular MSD and for the TAMSD are also explained in more detail. Finally, we present some data on the effect of the outlet width on the unclogging distributions and arch size statistics.
%
%

\end{abstract}

\maketitle


\paragraph{Simulation Geometry and Protocol}

The numerical simulations were carried out using the `granular' package in the molecular dynamics software LAMMPS (http://lammps.sandia.gov/)~\cite{Plimpton1995}. The computational domain is a quasi-2D hopper as shown in Fig.~1(a) of the main article.  Although the particles are spherical,  the hopper is effectively two-dimensional since after each time step the force and velocity on each grain in the third dimension is set to zero.
The side walls are created using `wall/gran' in LAMMPS. The inclined walls at the bottom of the hopper consist of highly overlapping spheres that are fixed in place to create a smooth line (Fig. 1(b) of the main article). 

The grain-grain and grain-wall forces are modeled by a Hertzian contact law coupled with a velocity dependent damping term. The mass of the small sphere is $M = 1.0$ and all spheres have the same density. The force between two overlapping spheres of radii $R_i$ and $R_j$ and mass $M_i$ and $M_j$ is a function of their overlap $\delta = R_i + R_j - r$, with $r$ the center to center distance between the two adjacent grains:
\begin{equation}
  F_n = \sqrt{\delta}\sqrt{\frac{R_i R_j}{R_i + R_j}}[k_n \delta {\bf n_\text{ij}} - m_\text{eff} \gamma_n {\bf v_n}]
\end{equation}

\begin{equation}
  F_t = \sqrt{\delta}\sqrt{\frac{R_i R_j}{R_i + R_j}}[k_t {\bf u_\text{ij}} - m_\text{eff} \gamma_t {\bf v_t}]
\end{equation}
$F_n$ and $F_t$ are the normal and tangential force components between the two grains with corresponding elastic constants $k_n = 2\times10^5$ and $k_t = \frac{2}{7}k_n$~\cite{Silbert2001}. The first term in the normal force is along ${\bf n_\text{ij}}$, a unit vector along the line connecting the centers of the contacting spheres. ${\bf u_\text{ij}}$ is the tangential displacement vector that is truncated to satisfy the Coulomb yield criterion ($F_t \leq \mu F_n$), $\mu$ being the static coefficient of friction. $m_\text{eff}=\frac{M_i M_j}{M_i + M_j}$ is the reduced mass. ${\bf v_n}$ and ${\bf v_t}$ are the normal and tangential components of the relative velocity between the two grains in contact.  The viscoelastic damping constant for normal contacts in our study is $\gamma_n = 500$, but we set $\gamma_t = 0$ for tangential contacts. The stiffness constants have been non-dimensionalized by $Mg/d_1^2$, and the damping constants by $1/d_1t$. The coefficient of static friction is $\mu_g$ = 0.8 for grain-grain interactions and $\mu_w$ = 0.5 for grain-wall interactions. 

At the start of simulation, 50000 grains with zero initial velocity are distributed uniformly between height 25 and 400 within the hopper. After removing grains with large overlaps, 1550-1650 grains remain in the hopper. These remaining, non-overlapping grains settle under gravity while the bottom of the hopper is kept closed. The height of grains in the hopper after settling is $\approx 100$. The grains settle until the total kinetic energy of the system is less than $10^{-12}$. Next, the hopper outlet is opened, and the grains fall out until an arch clogs the flow. To unclog this state, the lower inclined walls are vibrated vertically. For each realization of a clog, we measure the time from the initiation of the vibrations to the time when the center of any grain first exits the outlet. 

The range of  vibration strengths that were effective at unclogging the hopper were $ \Gamma = 2.8\textendash14.0 $.  Even at the lowest value, $\Gamma = 2.8$, there were arches that did not break for the maximum feasible simulation time, $ T_{sim} = \num{2e6} $.  At lower vibration strengths, therefore, no meaningful statistical analysis of  arch failure would have been possible. In experiments~\cite{zuriguel2014clogging,lozano2015stability}, arches fail at much smaller vibration strengths: $\Gamma < 2.5$.   We attribute the enhanced stability of our arches to the spheres being much ``softer'' than steel or glass grains, and to the large damping constants that are required  
to achieve mechanical equilibrium within simulation times.

\begin{figure}[h]
\begin{subfigure}[t]{0.4925\linewidth}
\centering
\resizebox{\linewidth}{!}{\includegraphics{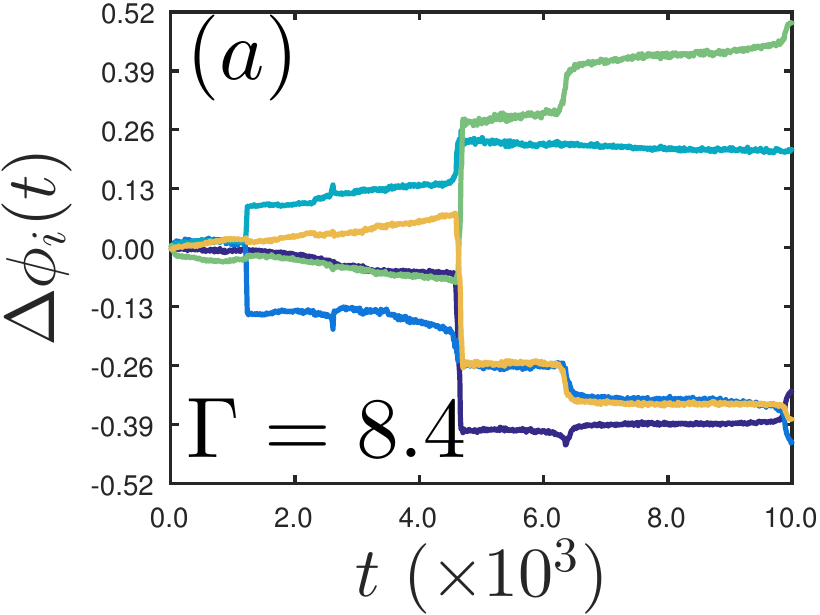}}
\end{subfigure}
\begin{subfigure}[t]{0.4925\linewidth}
\centering
\resizebox{\linewidth}{!}{\includegraphics{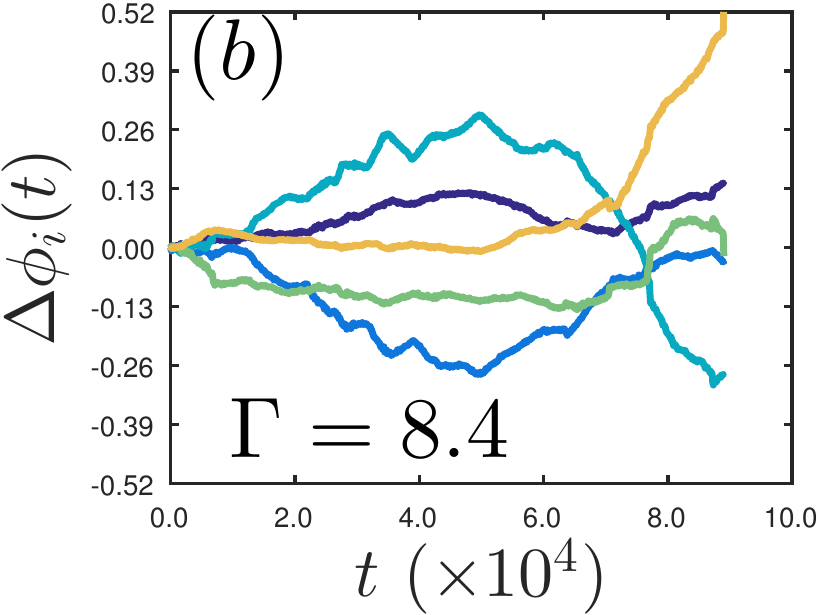}}
\end{subfigure}
\begin{subfigure}[b]{0.4925\linewidth}
\centering
\resizebox{\linewidth}{!}{\includegraphics{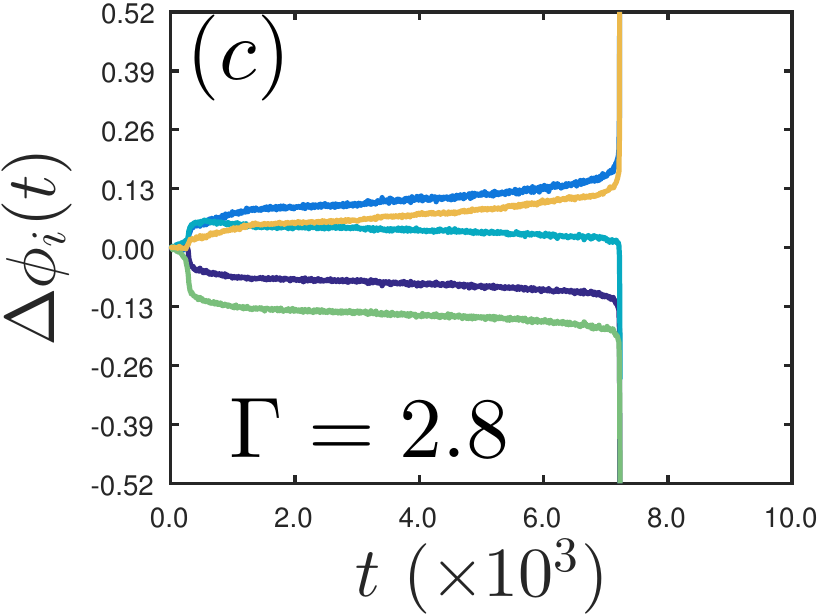}}
\end{subfigure}
\begin{subfigure}[b]{0.4925\linewidth}
\centering
\resizebox{\linewidth}{!}{\includegraphics{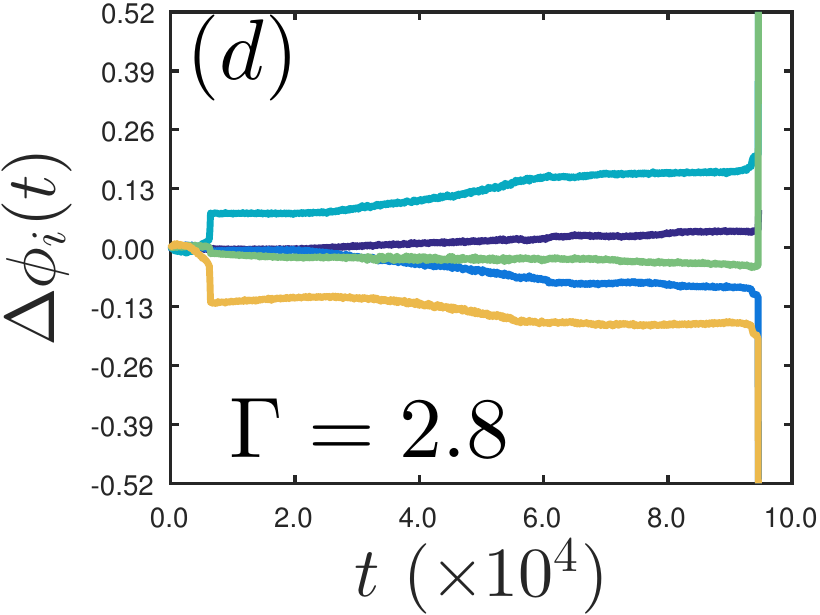}}
\end{subfigure}
\caption{\label{fig1} Time series of changes to the arch opening angles for four different five-angle arches unclogged using two amplitudes, $ \Gamma = 8.4 $ (top panel) and $ \Gamma = 2.8 $ (bottom panel). $ \Delta \phi(t) $ indicates the difference in radians from the initial value of the angle in the clogging arch. The colors indicate different angles within a single arch.  Arches can exhibit a series of reconfiguration events, when all the angles jump by a large amount. Plots $ (a) $ and $ (c) $ show intermediate unclogging times, $ t = 10,156 $ and $ t = 7,234 $, while $(b)$ and $(c)$ show long lived arches with lifetimes $ t = 89,048 $ and $ t = 94,607 $.}
\end{figure}

\paragraph{Examples of Arch Angle Dynamics}

We have included four new panels here in addition to Fig. 2 of the main article to highlight the changes in the trajectories from one realization of a five-angle arch to the next. Note the difference in the time scales between the arches lasting about $ t \approx 10^4 $ in $(a)$ and $(c)$, and the arches lasting $ t \approx 10^5 $ in $(b) $ and $(d)$. These arch opening angle time series suggest that, like the unclogging time distributions, the waiting times between reconfiguration events also follow a broad distribution. We support this qualitative observation through our analysis of the MSD and TAMSD for ensembles of many arches.

\paragraph{Ensemble MSD: Amplitude Comparison}

\begin{figure}[b]
\begin{subfigure}[t]{\linewidth}
\centering
\resizebox{!}{6 cm}{\includegraphics{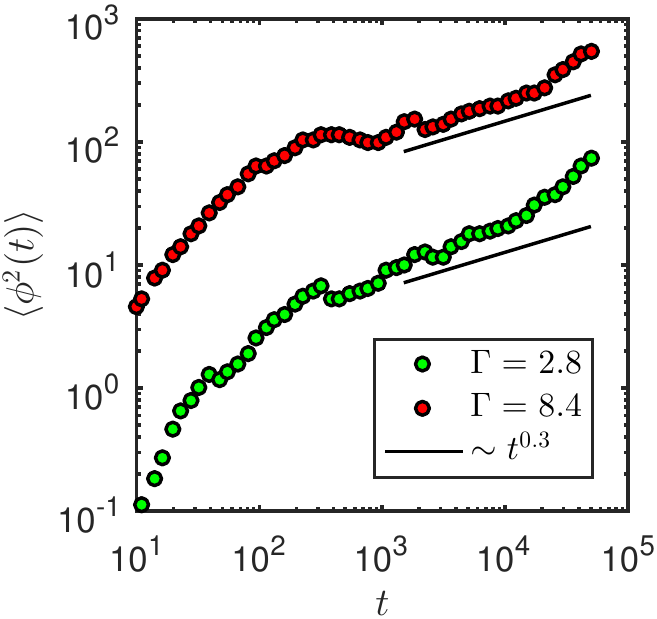}}
\end{subfigure}
\caption{\label{fig2} (color online) Time evolution of the ensemble-averaged MSD $ \langle \phi^2(t) \rangle $ for two different sets of five-angle arches unclogged using two amplitudes, $ \Gamma = 5.4 $ and $ \Gamma = 2.8 $. The black lines show a comparison to the expected slope $ \sim t^{0.3} $ based on the TAMSD analysis in the main paper. The lines are shown for the range of times analyzed, $ T = 1500\textendash50000$. }
\end{figure}

The usual way of identifying and quantifying anomalous diffusion is by studying the  ensemble-averaged mean-squared displacement (MSD), without any time averaging. For sets of arches of the same size that are unclogged at the same amplitude, the MSD is calculated as $ \langle \phi^2(t) \rangle = \frac{1}{N(t)}\sum_{i = 1}^{i=N(t)} \Delta \vec{\phi}_i(t)\cdot \Delta \vec{\phi}_i(t) $, where the index $ i $ indicates individual arches. The number $ N(t) $ of arches available for the ensemble average diminishes as more and more arches in the ensemble break.  

Anomalous diffusion is characterised by an MSD of the form $ \langle \phi^2(t) \rangle = D_\alpha t^\alpha $, where $ D_{\alpha} $ is a generalized diffusion coefficient, and an exponent  $ \alpha < 1 $ indicates subdiffusion.   For both $ \Gamma = 2.8 $ and $ \Gamma = 8.4 $, Fig. \ref{fig2} shows that the MSD grows sub-diffusively, with an exponent  $ \alpha = 0.3 $ consistent with the  one obtained from the ensemble-averaged TAMSD scaling (main text Fig. 3 (e) and (f))). The magnitude of the MSD for the smaller amplitude is, however, always a factor of $ \sim 10 $ less than the MSD for the larger amplitude. This indicates a corresponding reduction of the magnitude of the effective diffusion coefficient $ D_\alpha $. Finally, a steeper, short time increase of the MSD can be seen in the first $ 100 $ vibrations. This faster growth is probably because an initial readjustment or settling of the arch shape often happens right after the vibrations begin. The arches that break during this initial time span are the ones contributing to the peak in the unclogging time distributions at short times (see Fig. 5 in main text), whereas the arches which survive much longer follow the CTRW-like dynamics. These are the arches that lead to the broad distribution of unclogging time.

\paragraph{TAMSD: Averaging Protocol}

The TAMSD analysis is conducted for ensembles of arches with  $ N_\phi = 5 $ opening angles. We choose to restrict the analysis to these arches because they are the most common among the clogging arches occurring at the opening size studied, $ L = 4.2 $ (see Fig. \ref{fig4}(c)).  Analysis of smaller ensembles of arches of size $ N_\phi = 3 $ showed the same qualitative features.

To obtain the exponent $\alpha$, we want to study the scaling behavior of the TAMSDs for $ (\Delta << T) $, so as a practical cutoff, all the calculations are restricted to $ \Delta < 0.5*T $. Further, since we are interested in characterizing the dynamics of the arches \textit{prior} to the failure, the measurement time $ T $ extends only up to $ 100 $ vibrations before the recorded unclogging time for each arch. This restriction was found to be sufficient to ensure that the divergence of the arch angles during the final buckling event did not influence  the time averaging (see Fig.~\ref{fig1}). 

Once the TAMSD for each individual arch has been calculated for the available ranges of $ \Delta $ and $ T $, the ensemble averaged TAMSD $ \langle \delta^2(\Delta, T) \rangle $ is computed in the same way as for the regular MSD: $ \langle \delta^2(\Delta, T) \rangle  = \frac{1}{N(T)}\sum_{i = 1}^{i = N(T)} \delta_i^2(\Delta, T)$, where the index $i$ indicates individual arches. The number $ N(T) $ of unbroken arches diminishes with the overall averaging time $T$, decreasing  faster with  larger vibration amplitudes. For the $ \Gamma = 2.8 $, the ensemble of five-angle arches starts with  $ N(T=1500) = 704 $ but ends with  $ N(T=50000) = 303 $. For  $ \Gamma = 8.4 $, the initial ensemble size is $ N(T = 1500) = 472 $, while the final ensemble size is $ N(T = 50000) = 67 $, which was the longest time recorded. The ensembles of arches used in the analysis of the arch dynamics were distinct from the ensembles of arches used to estimate the unclogging time distributions. For the former, we recorded the full time series of positions for grains in and around the arch upto a maximum 100,000 vibrations. For the unclogging time distributions, we only kept data on the initial state and on the final unclogging times.     

\begin{figure}[t]
\begin{subfigure}[t]{0.4925\linewidth}
\centering
\resizebox{\linewidth}{!}{\includegraphics{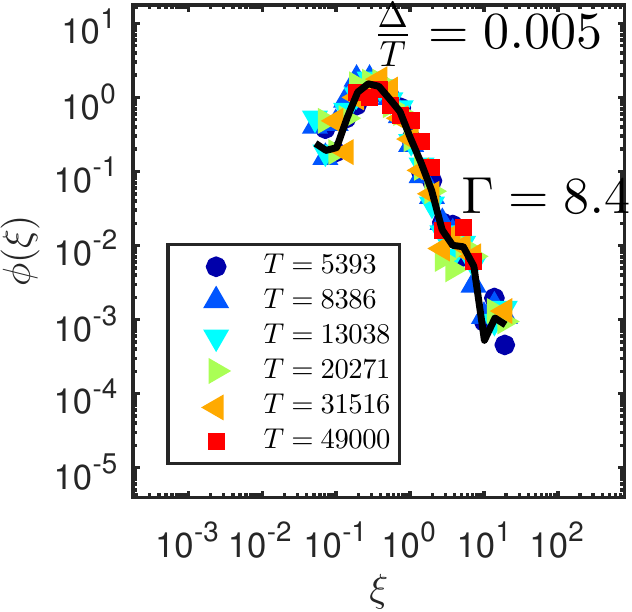}}
\end{subfigure}
\begin{subfigure}[t]{0.4925\linewidth}
\centering
\resizebox{\linewidth}{!}{\includegraphics{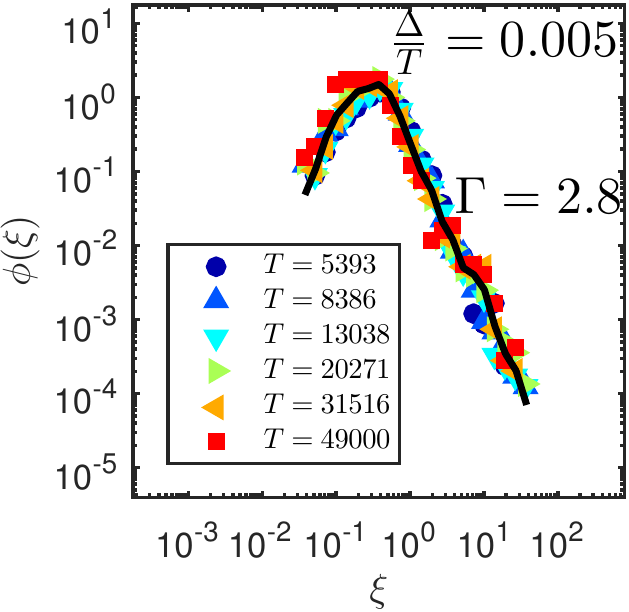}}
\end{subfigure}
\begin{subfigure}[b]{0.4925\linewidth}
\centering
\resizebox{\linewidth}{!}{\includegraphics{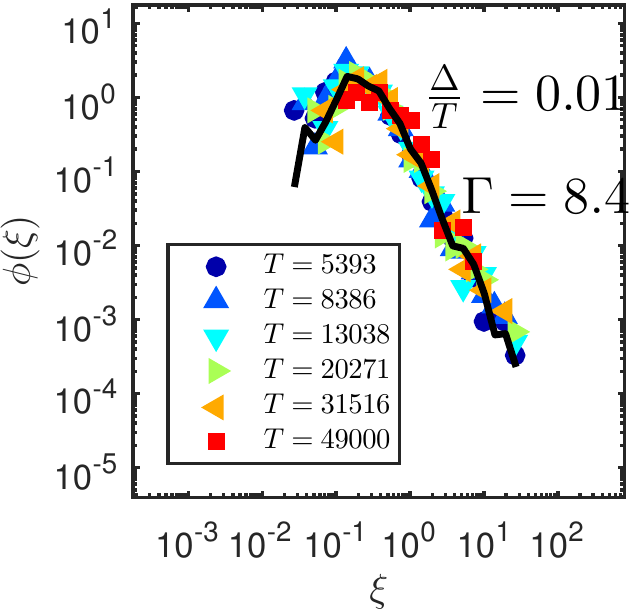}}
\end{subfigure}
\begin{subfigure}[b]{0.4925\linewidth}
\centering
\resizebox{\linewidth}{!}{\includegraphics{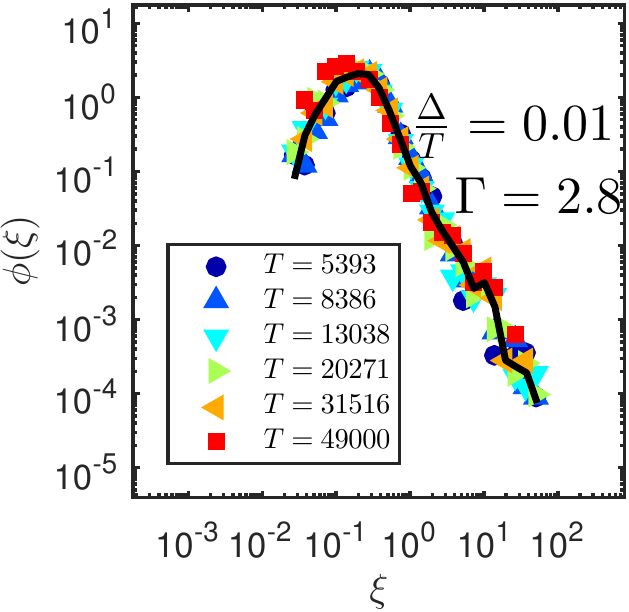}}
\end{subfigure}
\begin{subfigure}[b]{0.4925\linewidth}
\centering
\resizebox{\linewidth}{!}{\includegraphics{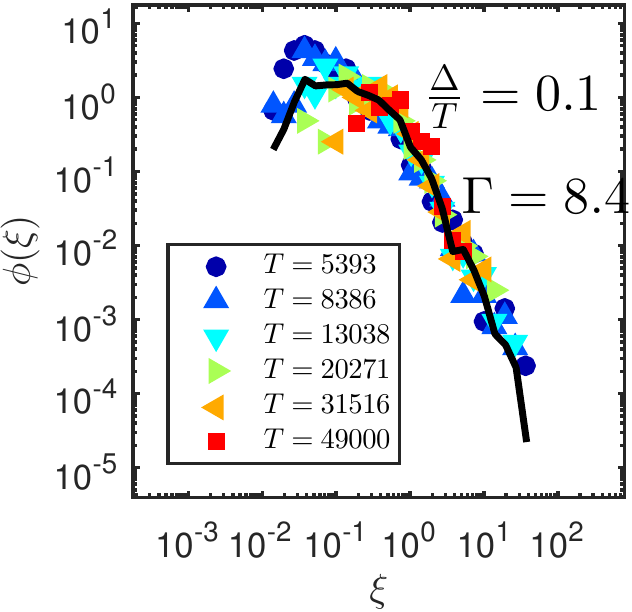}}
\end{subfigure}
\begin{subfigure}[b]{0.4925\linewidth}
\centering
\resizebox{\linewidth}{!}{\includegraphics{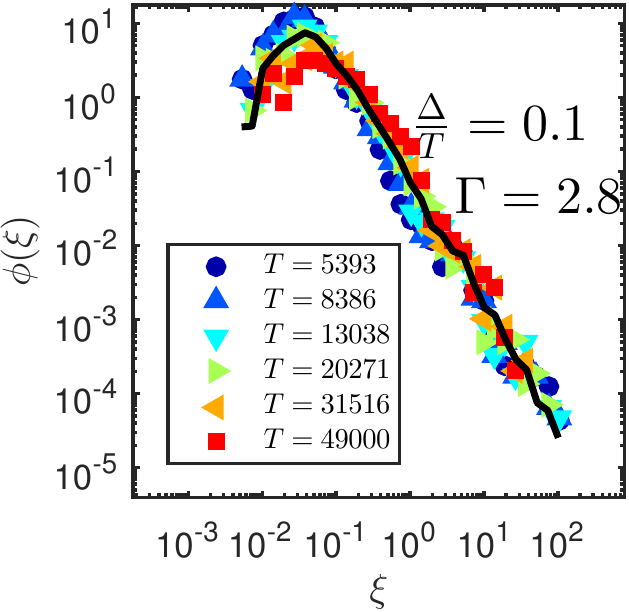}}
\end{subfigure}
\caption{\label{fig3} Distributions $ \phi(\xi) $ of the normalized TAMSD for five-angle arches unclogged with amplitudes, $ \Gamma = 8.4 $ (left column) and $ \Gamma = 2.8 $ (right column). The distributions are grouped by the ratio $ \Delta/T $ to show that for the range of times observed, the distribution shape is mainly controlled by this ratio. The legends give the corresponding values of the overall measurement time $ T $ for each colored distribution. The solid lines are the averaged $ \phi(\xi) $ at each value of $ \Delta/T $ which were used to compute the averaged value of the EB parameter $ \langle \xi^2 \rangle - \langle \xi \rangle^2  $ as a function of $ \Delta/T $ as shown in Fig. 4 of the main article.}
\end{figure}

\paragraph{Shape/Behavior of TAMSD Distributions}

The distributions $ \phi(\xi) $ of the normalized TAMSD variable $ \xi = \delta^2/\langle \delta^2 \rangle $ are found to be well-described as a function of the ratio $ \Delta/T $ over the range of data available. The distributions are shown grouped by $ \Delta/T $ in the panels of Fig.~\ref{fig3} for $ \Gamma = 2.8 $ (left column) and $ \Gamma = 8.4 $ (right column). The narrow distributions at small $ \Delta/T $ occur in the region where the ensemble TAMSD appears flat (see Fig.~3 main text). We believe this region indicates a short time dynamics where the CTRW scaling does not yet apply. There is a comparable short time scale seen in the regular MSD (Fig.~\ref{fig2}). In the region where the growth of the TAMSD is linear, there is a range of $ \Delta/T $ where the distributions $ \phi(\xi) $ appear to acquire a constant form, independent of $ \Delta/T$. (However, as $ T $ increases, the statistics available for the time-averaging are diminished, especially for $ \Gamma = 8.4 $). In this range, $ \Delta/T = 0.1\textendash0.4$, the EB parameter, the variance of the TAMSD distribution, reaches a roughly constant value at both amplitudes. Therefore, we claim that the distributions shown for $ \Delta/T = 0.1 $ give the best characterization of the random TAMSD variable. These distributions are extremely broad compared to the $ \phi(\xi) $ distributions one would predict for a CTRW with a power law waiting time distribution~\cite{He2008b}. 

\paragraph{Effect of Outlet Width on Unclogging}

\begin{figure}
\begin{subfigure}[t]{0.49\linewidth}
\centering
\resizebox{\linewidth}{!}{\includegraphics{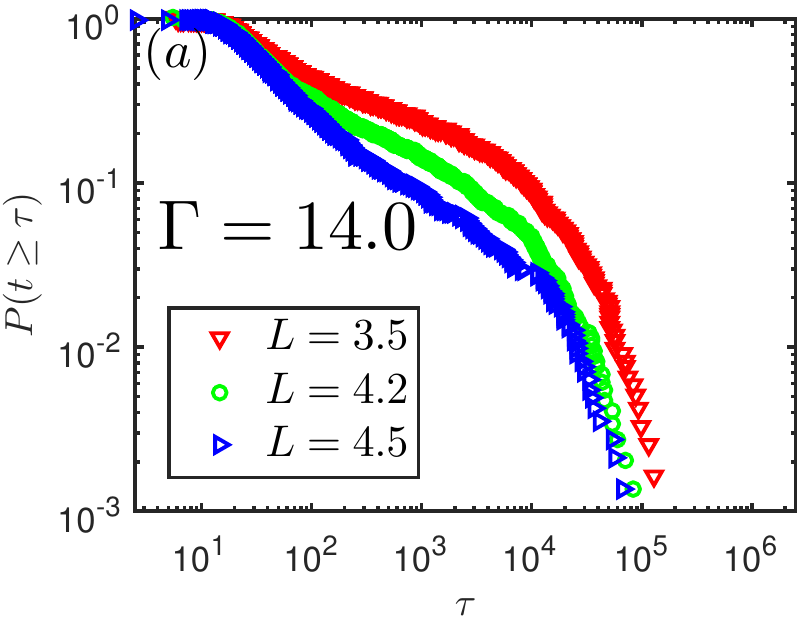}}
\end{subfigure}
\begin{subfigure}[t]{0.49\linewidth}
\centering
\resizebox{\linewidth}{!}{\includegraphics{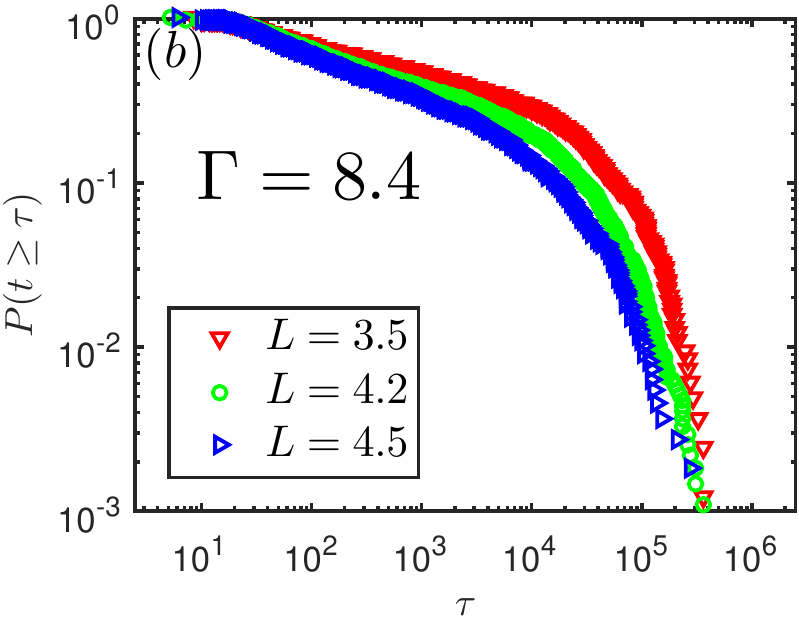}}
\end{subfigure}
\begin{subfigure}[t]{\linewidth}
\centering
\resizebox{!}{4 cm}{\includegraphics{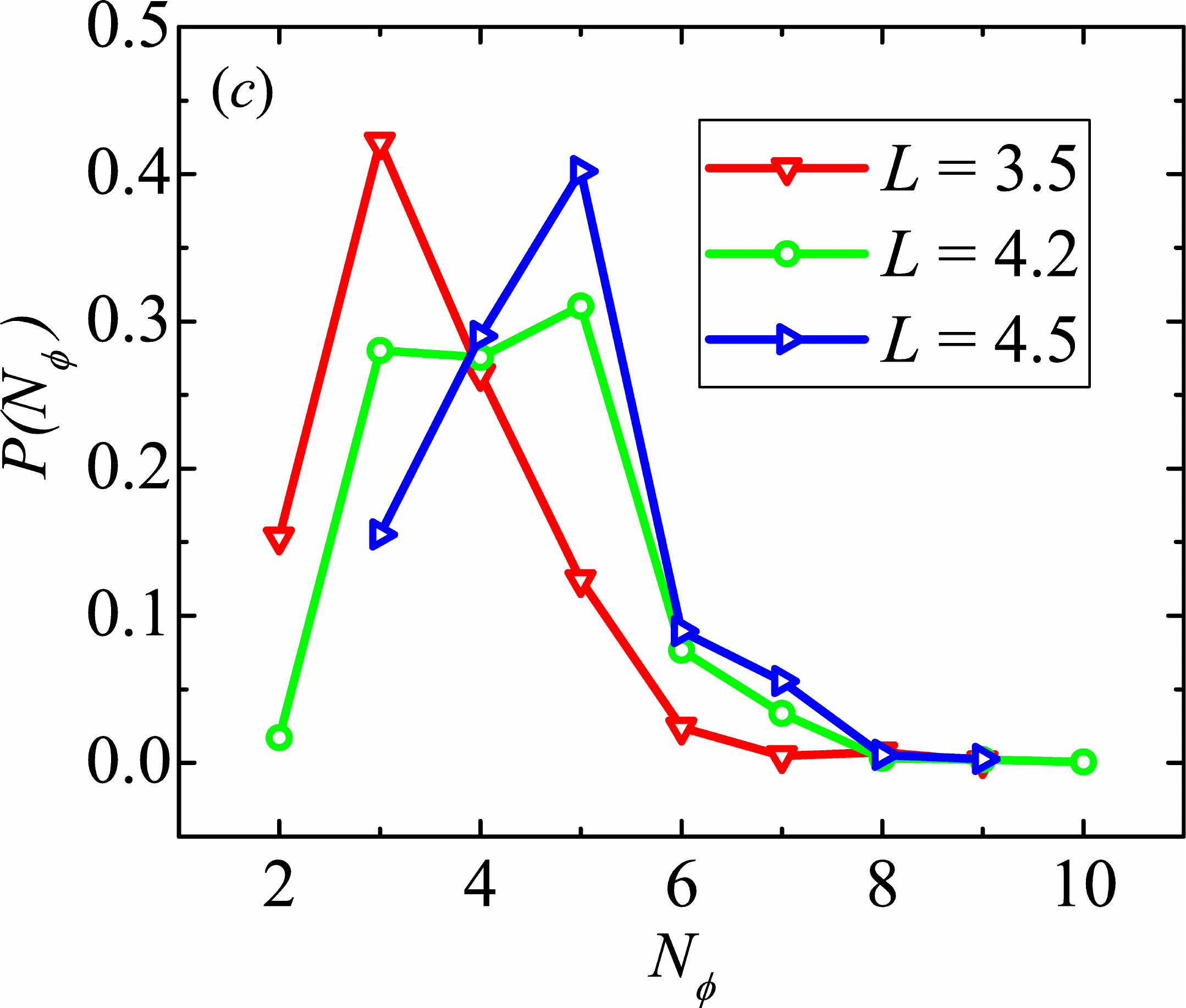}}
\end{subfigure}
\caption{\label{fig4} Variation of the unclogging times CCDF with outlet width $ L $ for $ (a) $ amplitude $ \Gamma = 14.0 $ and $(b)$ amplitude $ \Gamma = 8.4 $. For our hopper geometry, the dependence of the unclogging time distributions on the opening size is found to be much less pronounced than the dependence on the vibration amplitude. The three widths shown are $ L = 3.5, 4.2, 4.5 $. For $ \Gamma = 14.0 $, the ensembles contain $ N = 1201,1469, {\rm and}~1441 $ arches. For $ \Gamma = 8.4 $, the ensembles contain $ N = 830, 2718,{\rm and}~1097 $ arches. A histogram of the arch sizes for these three widths is shown in (c).}
\end{figure}

The opening width, $L$, of the hopper was changed by keeping the angle of inclined walls fixed at 45$^\circ$ to horizontal. We tested the dependence of the unclogging time distributions on the outlet size for three different outlet sizes $ L = 3.5, 4.2, {\rm and}~4.5 $, and two different vibration amplitudes, $ \Gamma = 14.0 $ and $ \Gamma = 8.4 $.  As seen in Fig. \ref{fig4}, reducing either $L$ or $\Gamma$  broadens the distribution, however,  the effect of changing $L$ is much weaker than that of changing $\Gamma$. 

Relating the unclogging time distributions to the picture of the arch dynamics as a CTRW through a series of stable arch configurations, we speculate that changing $ L $ influences the unclogging time distribution by controlling the distribution of arch sizes present at the outlet opening.  We expect that changing the arch size distribution should change the size of the effective phase space of stable arch configurations available for the arch to explore before failure, which in turn should influence the shape of the unclogging time distributions. As seen in Fig. \ref{fig4}, the distributions of arch sizes change only slightly for the range of opening sizes $ L $ that were tested. This weak dependence of the arch size distribution on $ L $ may be due to the tapered geometry of the hopper in our simulations:  since arches can form over a range of heights in the tapered region the precise value of $ L $ becomes less important. For hoppers with flat walls adjacent to the outlet opening, the dependence of the arch sizes distributions, and hence of the unclogging time distributions, might depend more sensitively on $ L $. 

\bibliographystyle{apsrev4-1}

%